\documentclass[epj,draft,nopacs]{svjour}
\def\journal#1, #2, #3#4, #5#6#7#8    {
    {#1~} {#2}  (#5#6#7#8) #3#4}

\def\prd{\journal Phys. Rev. D, }

\def\rmp{\journal Rev. Mod. Phys., }

\def\npb{\journal Nucl. Phys. B, }
\def\plb{\journal Phys. Lett. B, }

\def\ijmpa{\journal Int. Jour. Mod. Phys. A, }

\def\jmp{\journal J. Math. Phys., }
\def\jhep{\journal J. High Energy Phys., }

\def\Rb{{I\!\! R}}
\def\Ii{{1\!\! {\rm I}}}

\newcommand{\beq}[1]{\begin{equation}\label{#1}}
\newcommand\eeq{\end{equation}}
\newcommand{\ba}[1]{\begin{eqnarray}\label{#1}}
\newcommand{\baa}{\begin{eqnarray}}
\newcommand\ea{\end{eqnarray}}
\newcommand{\bee}{\begin{equation}}

\def\nn{\nonumber \\}

\def\o{\omega}

\newcommand{\h}{Hamiltonian}

\def\hlf{\frac{1}{2}}

\newcommand{\nc}{noncommutative}

\begin{document}

\title{Representations of noncommutative  quantum mechanics and symmetries}

\author{Larisa Jonke\thanks{larisa@irb.hr} \and Stjepan Meljanac\thanks{meljanac@irb.hr}}

\institute{Theoretical Physics Division,\\
Rudjer Bo\v skovi\'c Institute, P.O. Box 180,\\
HR-10002 Zagreb, Croatia}

\date{Received: date / Revised version: date}

\abstract{
We present a unified approach to representations of quantum mechanics
on noncommutative  spaces with general constant commutators of
phase-space variables.
We find two phases
and  duality relations among them in arbitrary dimensions. 
Conditions for physical equivalence of different representations of 
a given system are analysed. Symmetries and classification of phase spaces
are discussed. 
Especially, the dynamical symmetry of a physical system 
is investigated.
Finally, we apply our analyses to the two-dimesional harmonic 
oscillator and the Landau problem.
}

\maketitle


\section{Introduction}

The problem of quantum mechanics  on \nc \ spaces can be 
understood in the framework of deformation quantization.
It is a subject with a long history starting with works of Wigner, Weyl and
von Neumann (see Refs.\cite{old}  for a recent review).
More recently, the investigation of \nc \ quantum mechanics was
inspired  by the development that led to 
\nc \ field theory.
Namely, it was realized 
that low-energy effective field theory of various 
D-brane configurations has a configuration space  which is described in 
terms of noncommuting, matrix-valued coordinate fields \cite{witt}.
Then, it was shown that, in a certain limit, the  entire string 
dynamics can be  described by minimally 
coupled gauge theory  on noncommutative space \cite{sw}.
Intensive studies of  field theories on various noncommutative spaces 
\cite{rev}
were also inspired by connection with M-theory compactifications  \cite{sch}
and more recently, by the matrix formulation of the 
quantum Hall effect \cite{qhe}.
In order to study phenomenological consequences of noncommutativity, 
a \nc \ deformation of Standard model have been constructed and
analysed \cite{w}.

In the last two years a lot of work has been done in analysing and 
understanding quantum mechanics 
(QM) on noncommutative (NC) spaces 
\cite{nair,nairp,jell,torus,bell,acatr,smyr,sma,esp,lm,gamboa}
and also in applying it to different physical systems
in order to test its relevance to the real world \cite{gam,ab}. 
Still, there are many 
different views and approaches to  \nc \ physics \cite{ph,rb,ad}. 
Some important questions, 
such as 
physical equivalence of different \nc \ systems, as well as their relation to 
ordinary quantum mechanics with canonical variables have  not been completely 
resolved.
The symmetries and the
physical content in different phases have  not been completely elucidated,
even in the simplest case of harmonic oscillator on the \nc \ plane.

In this paper, we present a unified approach to representations of 
NCQM in arbitrary dimensions. 
Conditions for physical equivalence of different representations of
a given system are analysed. We show that there exist two phases in 
parameter space.
Phase I can be viewed as a smooth deformation of ordinary QM, 
where all physical quantities have a smooth limit to physical 
quantities in ordinary QM.
Phase II is qualitatively different from 
phase I and cannot be continuously connected to ordinary QM.
There is a discrete duality transformation connecting the two phases.

Futhermore, we investigate symmetry transformations
preserving commutators, the \h \
and also the dynamical symmetry of the physical system. 
We analyse
the angular momentum generators, and give conditions 
for their existence.

We demonstrate our general results on the simple example of 
harmonic oscillator on a \nc \ plane.
Especially, we describe dynamical 
symmetry structure and discuss uncertainty  relations.
Finally, we briefly comment
on the NC Landau problem.

\section{Noncommutative quantum mechanics  and its representations}

Let  us start with the two-dimensional  \nc \ coordinate plane $X_1,X_2$ and 
the corresponding momenta $P_1,P_2$, where $X_i$ and $P_i$ are 
hermitean operators. We describe a problem in 4-dimensional phase space
using variables
$U=\{U_1,U_2,U_3,U_4\}=\{X_1,P_1,X_2,P_2\}$, where  $U_i$'s satisfy 
general commutation relations
\bee
[U_i,U_j]=iM_{ij},\;i,j=1,2,3,4,\label{m1}\eeq
and $M_{ij}=-M_{ji}$ are real constants (c-numbers). 
The antisymmetric matrix $M$ is 
parametrized by 6 real parameters
\baa\label{m2}
M=\left(\begin{array}{cccc}
 0&\hbar_1&\theta&\phi_1\\
-\hbar_1&0&\phi_2&B\\
-\theta&-\phi_2&0&\hbar_2\\
-\phi_1&-B&-\hbar_2&0\end{array}\right)\ea
and the 
determinant $\det M=(\hbar_1\hbar_2-\theta B +\phi_1\phi_2)^2$ is 
positive. The critical point \mbox{$\det M=0$} divides the space 
of the  parameters into two phases: phase I for $\kappa=\hbar_1\hbar_2-\theta B 
+\phi_1\phi_2>0$ and phase II for 
$\kappa<0$. 
The ordinary, commutative space $M_0$ 
\baa\label{comm}
M_0=\left(\begin{array}{cccc}
          0& 1& 0& 0 \\
             -1&0&0&0 \\
                0&0&0&1 \\
                0&0&-1&0 \end{array}\right)\ea
has $\kappa=1$ and 
belongs to phase I.
Therefore, we can view phase I as a continuous, smooth deformation
of ordinary quantum mechanics.
The critical point $\kappa=0$ corresponds to reduction of dimensions 
in phase space and to infinite degeneracy of states and is 
related to the (\nc ) Landau problem \cite{nairp} 
(see also the "exotic"
approach \cite{ph}).

If we define the  angular momentum $J$ as
\bee\label{defJ}
\left[ J,X_a\right] = i\varepsilon_{ab}X_b,\;\;
\left[ J,P_a \right] = i\varepsilon_{ab}P_b,\;a,b=1,2,\nonumber\eeq
or
\bee\label{def2J}
[J,U_i]=iE_{ij}U_j,\;i,j=1,2,3,4,\eeq
then for a given regular matrix $M$ we can construct the angular momentum 
only if
\mbox{$[E,M]=0$}.
This condition is fulfilled when $\phi_1=\phi_2$ and $\hbar_1=\hbar_2$.
Then
\bee\label{J}
J=-\frac{1}{2}(EM^{-1})_{ij}U_iU_j.\eeq
We see that for general $M$ the angular momentum in the usual sense
may not exist. Moreover, even when it exists, it may have unusual
properties. Namely, it was shown in Ref.\cite{bell} that in phase I system 
could have an infinite number of states for a given value of the 
angular
momentum, while in phase II the number of such states is finite.

Now, let us assume that the \h \ of the system describes the  motion of a 
single particle on a \nc \ plane:
\bee\label{h1}
H=\frac{1}{2}{\bf P}^2+V({\bf X}^2) , \eeq
with a discrete spectrum $E_{n_1,n_2}$, 
where $n_1,n_2$ are nonnegative integers. 
The pair $(H(U),M)$ defines a system with a given energy spectrum and the 
corresponding energy eigenfunctions. We wish to characterize  all systems
$(H'(U'),M')$ 
with the same spectrum. The class of such systems is very large and 
can be defined by all real, nonlinear, regular transformations
$U'_i=U'_i(U_j),\;U_i=U_i(U'_j)$. We restrict ourselves to linear 
transformations $Gl(4,\Rb)$ in order to keep the 
matrix elements $M'_{ij}$ independent of phase-space variables.
Among these, of special interest are $O(4)$ orthogonal transformations
changing  commutation relations, and 
the group of
transformations isomorphic to $Sp(4)$
keeping
$M$ invariant. Systems with the same energy spectrum 
connected by transformations that keep commutation 
relations invariant are physically equivalent. 
In both cases, the  \h \ generally changes, but the energy 
spectrum is 
invariant.

Let us consider $O(4)$ transformations. 
The important property \cite{bk,smyr} is 
that there exists an orthogonal transformation $\tilde R$ such that
\baa\label{r1}
 \tilde R^T M \tilde R =\left(\begin{array}{cccc}
          0& |\o_1|& 0& 0 \\
                  -|\o_1|&0&0&0 \\
                0&0&0&|\o_2| \\
                0&0&-|\o_2|&0 \end{array}\right),\ea
where $|\o_1|\geq |\o_2|\geq 0$ and $\det M=
\o_1^2\o_2^2\geq 0$. The matrix $\tilde R$ is unique up to the transformations 
$S\in SO(4)$: 
\bee\label{sim}
 \tilde R_s=S\tilde R  ,\;\;S^T M S=M.\eeq
The first (second) phase is characterized by ${\rm det }\tilde R=+1$
$({\rm det } \tilde R = -1 )$.
For  the two-dimensional case,  
the  eigenvalues of the general matrix $iM$, Eq.(\ref{m2}),
are 
\baa\label{op}
\o_{1,2}&=&\frac{1}{2}\sqrt{(\theta-B)^2+(\phi_1+\phi_2)^2+(\hbar_1+\hbar_2
)^2}\nn
&\pm &\frac{1}{2}\sqrt{(\theta+B)^2+(\phi_1-\phi_2)^2+(\hbar_1-\hbar_2)^2}.\ea
Notice  that $\o_1$ is always positive, while $\o_2$ changes the sign at the 
critical point $\det M =0$, i.e., when
$\theta B-\phi_1\phi_2=\hbar_1\hbar_2.$

The matrix $\tilde R$ is universal, i.e., there exists the matrix $R\in SO(4)$, 
with $\det R=1$ such that
\baa\label{r2}
 R^T M R =\left(\begin{array}{cccc}
          0& \o_1& 0& 0 \\
                  -\o_1&0&0&0 \\
                0&0&0&\o_2 \\
                0&0&-\o_2&0 \end{array}\right)\equiv J_{\o},\ea
regardless of  $\o_2$ being positive, zero, or negative. When $\o_2<0$, we use 
$\tilde R= RF$ to obtain Eq.(\ref{r1}), where the 
flip matrix $F\in O(4)$ is given by
\baa\label{f}
F=\left(\begin{array}{cccc}
          1& 0& 0& 0 \\
             0&1&0&0 \\
                0&0&0&1 \\
                0&0&1&0 \end{array}\right).\ea
At the critical point $(\o_2=0)$ both $R$ and $RF$ satisfy Eq.(\ref{r1}).

The most general orthogonal matrix $R$ depends on six continuous 
parameters. For a fixed values $\{\o_1,\o_2\}$, the number of 
parameters of the matrix $M$ is the same as the  number of 
parameters of $R$.
As we have already mentioned, there exist orthogonal 
matrices that commute with $M$, Eq.(\ref{sim}), and these matrices 
form a group isomorphic to $U(1)\times U(1)$.
We can use this symmetry to fix two parameters in the matrix $M$, 
and we choose  $\hbar_1=\hbar_2=1$ 
or  $\hbar_1=-\hbar_2=1$. This parametrization 
covers all  pairs $\{\o_1,\o_2\}$ such that $\o_1^2+\o_2^2\geq 2$.

The  eigenvalues $\o_1,\o_2$ have the meaning of the "Planck" constants for
new variables:
\baa\label{u0}
&& U_i^0=R^T_{ik}U_k,\nn
&&[X_a^0,P_b^0]=i\o_a\delta_{ab},\;[X_a^0,X_b^0]=[P_a^0,P_b^0]=0.\ea
\hfill{We have transformed the \nc \ system}\\
$(H(U),M)$ into $(H(RU^0),J_{\o})$ 
keeping 
the energy spectrum of the system invariant. 
Note that for  system $(H(RU^0)$, $J_{\o})$ we cannot 
define the angular momentum. In order to connect a \nc \ system 
with a quantum mechanical system in ordinary space,  we perform the 
following transformation:
\baa\label{diag}
U^0=Du^0=\left(\begin{array}{cccc}
          \sqrt{\o_1}& 0& 0& 0 \\
             0&\sqrt{\o_1}&0&0 \\
                0&0&\sqrt{|\o_2|}&0 \\
                0&0&0&\sqrt{|\o_2|} \end{array}\right)u^0,\ea
where the variables $u^0=\{u^0_1,u^0_2,u^0_3,u_4^0\}$ are canonical, i.e., 
$[x_a^0,p_b^0]=i\delta_{ab}$ and $[x_a^0,x_b^0]=[p_a^0,p_b^0]=0$.
Now we have obtained $H(U)=H(RDu_0)$ with the standard canonical relations 
$M_0$, Eq.(\ref{comm}). 
Transformation $D$, Eq.(\ref{diag}) 
is valid in both phases, but  at the critical 
point it becomes singular. 
Also note that the composition ${\cal L}_0=R D$ has a smooth limit when 
$M\rightarrow M_0$.

In order to make contact with other representations in the literature 
\cite{nairp,sma}, we 
perform a symplectic transformation on the canonical variables $u_i^0$:
\bee\label{s}
u_i=S_{ij}u_j^0=V u_i^0 V^{\dagger},\eeq
where $S$ commutes with $M_0$ and 
$$V=\exp{(i\sum v_{ij} u_i^0u_j^0)},\; \; VV^{\dagger}=1,$$
is a unitary operator corresponding to the symplectic transformation $S$.
This symplectic transformation generates a class of ordinary quantum 
mechanical systems which are physically (unitary) equivalent.
Of course, the initial system $(H(U),M)$ is not  physically equivalent to 
the canonical ones, but all corresponding physical quantities can be 
uniquely determined.
In Fig. 1 we show a simple graphic description of connection between 
different representations of NC quantum mechanics.

\setlength{\unitlength}{1cm}
\begin{center}
\begin{picture}(12,6)
\put(2,5.65){$\kappa>0$}
\put(7,5.65){$\kappa<0$}

\put(0.5,1.5){\vector(0,1){3}}
\put(3.5,4.5){\vector(-1,0){3}}
\put(3.5,1.5){\vector(0,1){3}}
\put(3.5,1.5){\vector(-1,0){3}}
\put(3.5,1.5){\vector(-1,1){2}}
\put(1.5,3.5){\line(-1,1){1}}
\put(0,1){$(u,M_0)$}\put(1.5,1.75){$S$}\put(2.85,1){$(u^0,M_0)$}
\put(3.75,3){$D$}\put(2,3){${\cal L}_0$}\put(0,3){$\cal L$}
\put(0,4.75){$(U,M)$}\put(2,4.15){$R$}\put(2.85,4.75){$(U^0,J_{\o})$}

\put(3.75,4.5){\vector(1,0){1.5}}
\put(5.25,4.5){\vector(-1,0){1.5}}
\put(4.5,4.15){$F$}
\put(3.75,1.5){\vector(1,0){1.5}}
\put(5.25,1.5){\vector(-1,0){1.5}}
\put(4.5,1.75){$F$}

\put(5.5,1.5){\vector(0,1){3}}
\put(5.5,4.5){\vector(+1,0){3}}
\put(8.5,1.5){\vector(0,1){3}}
\put(5.5,1.5){\vector(1,0){3}}
\put(5.5,1.5){\vector(+1,1){2}}
\put(7.5,3.5){\line(1,1){1}}
\put(5.0,1){$(Fu^0,FM_0F)$}\put(6.5,1.75){$FSF$}\put(7.5,1){$(Fu,FM_0F)$}
\put(4.85,3){$ D$}\put(6.45,3){$ {\cal L}'_0$}
\put(8.75,3){${\cal L}'$}
\put(5.0,4.75){$(FU^0,FJ_{\o}F)$}\put(6.8,4.15){$R'$}\put(8.0,4.75){$(U',M')$}

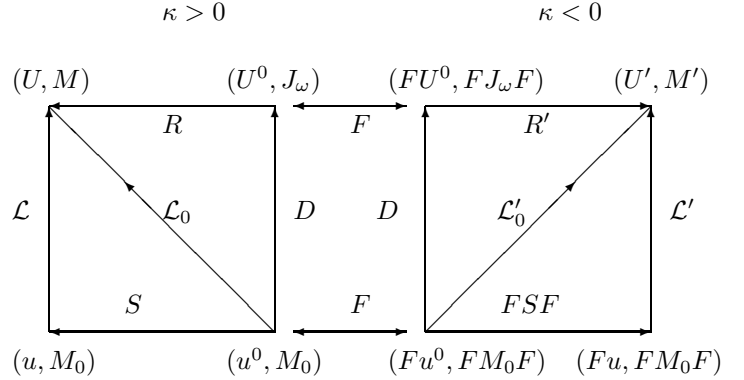
\put(0,0){Fig. 1: Graphic representation of the transformations.}

\end{picture}
\end{center}
There is a "mirror-symmetric" diagram for
phase II,
obtained using the flip matrix $F$ (\ref{f}), where 
$U'=R'FR^TU$,
$M'=R'FR^TMRFR'^T$, and ${\cal L}'=R'FR^T{\cal L}F$. 
The matrix $R'$ is any special orthogonal matrix. The universality 
of matrix $R$ means
that we can choose $R'$ and  $R$ to have the same functional dependence
on matrix elements $M'_{ij}$ and $M_{ij}$, respectively.
We have the discrete $Z_2$ symmetry
connecting two components of group $O(4)$, or more 
generally, $Gl(4,\Rb)$.

Starting from the matrix $M$, we can construct the matrix $R$ by finding 
eigenvalues and eigenvectors of the matrix $iM$, i.e.,
$R=U_M U_J^{\dagger} $, where
$$U_M^{\dagger}(i M)U_M=U_J^{\dagger}(iJ_{\o})U_J={\rm diag}(\o_1,-\o_1,\o_2,
-\o_2).$$
For example, for $\phi_1=\phi_2=0$, we can write the 
matrix $R$ in the following 
form:
\baa\label{rr}
R =\left(\begin{array}{cccc}
           \cos{\varphi}& 0& \sin{\varphi} & 0 \\
    0&\sin{\varphi} &0&\cos{\varphi} \\
     0&\cos{\varphi}&0& -\sin{\varphi}\\
     -\sin{\varphi}&0&\cos{\varphi}&0 \end{array}\right)\ea
where we choose $\varphi\in(0,\pi/2)$, $\theta\geq 0$, $\theta+B\geq 0$ and 
\bee\label{cs}
\cos{\varphi}=\frac{1}{\sqrt{1+(B+\o_2)^2}}=\frac{\o_2+\theta}
{\sqrt{1+(\o_2+\theta)^2}}=\sqrt{\frac{\o_1-B}{\o_1+\o_2}}.\eeq
The basic relations 
are
\baa
\o_1\o_2=1-\theta B,\nn\o_1-\o_2=\theta+B,\nn\o_1+\o_2=\sqrt{(\theta-B)^2+4}.
\nonumber\ea
An interesting  example of matrix $R$ is obtained in the case $\theta=B$, which 
corresponds to $\varphi=\pi/4$ in Eq.(\ref{rr}). In that case matrix $R$ 
does not depend on noncommutativity parameters.

The $R$ matrix was discussed in Ref.\cite{smyr} in the context of 
the 
$*$-genvalue problem, but only in phase I. The authors of Ref.\cite{smyr} 
stated that the matrix $R$ became singular at the  critical point. 
However, we 
wish to emphasize that the matrix $R$ is universal orthogonal matrix valid
in both phases and  even at the critical point.

The transformations ${\cal L}$, 
$S$  and ${\cal L}_0$ were  discussed in Refs.\cite{nairp,sma} for the case of 
two-dimensional harmonic oscillator and parametrization $\hbar_1=\hbar_2=1$ and 
$\phi_1=\phi_2=0$, with 
the identification $u^0=\{Q,P\}$ and $u=\{\alpha,\beta\}$. 
The authors of Ref.\cite{nairp} treated the two phases separately 
overlooking the universality of the transformation ${\cal L}_0=RD$, 
whereas in Ref.\cite{sma}
phase II was not analysed.

We point out that two systems $(H(U),M)$ and $(H'(U')$, $M')$ with the same 
energy spectrum and $M\neq M'$ are physically not equivalent. The condition
for physical equivalence is the same energy spectrum and the same 
commutation relations $M=M'$. Hence, even within the same phase two systems 
with the same energy spectrum can be quite different.

\section{Two phases, duality and symmetries   in arbitrary dimensions}

Construction of different representations of  quantum mechanics on 
a \nc\  plane can be easily generalized to arbitrary dimensions $D$.
The regular, antisymmetric matrix $M$ is
parametarized by $D(2D-1)$ real parameters.
We can classify \nc \ spaces according to
$\{\o_1,\o_2,\ldots,\o_D\}$, eigenvalues of the Hermitean matrix $iM$.
The determinant of the matrix 
$M$ is positive, $\det M=\o_1^2\cdots\o_D^2$.
The critical point $\det M=0$ divides the space of the parameters in two
phases. In phase I, $\kappa=\o_1\cdots \o_D>0$, and in phase
II, $\kappa<0$.
The critical point $\kappa=0$ may have interesting physical 
applications, like the Landau problem in two dimensions.

In $D$-dimensions, angular momentum operators are generators of coordinate
space rotations:
\baa\label{dj}
\left[J_{ab},X_c\right]&=&i(\delta_{ac}\delta_{bd}-\delta_{ad}\delta_{bc})X_d,\nn
\left[J_{ab},P_c\right]&=&i(\delta_{ac}\delta_{bd}-\delta_{ad}\delta_{bc})P_d,
\;a,b,c,d=1,\ldots,D\nonumber,\ea
and generally,
$$[J_{ab},U_i]=(E_{ab})_{ij}U_j,\;i,j=1,\ldots 2D.$$
For a regular matrix $M$ we can construct the angular momentum generators
$J_{ab}=-\frac{1}{2}(E_{ab}M^{-1})_{ij}U_iU_j$ 
only if
$[E_{ab},M]=0$, for all $a,b=1,\ldots,D$.

There are two sets of important transformations in the $2D$ phase space.
One is a group of linear transformations $U_i'=S_{ij}U_j$ preserving 
$M$, $S^TMS=M$. These transformations form a group $G(M)$ isomorphic to 
$Sp(2D)$. For every tranformation $S$ there exists a unitary 
operator $V\sim\exp{(i\sum v_{ij}U_iU_j)}$, and any two systems connected
by such an $S$ transformation are physically (unitary) equivalent.

The other important set of transformations are orthogonal transformations
$O(2D)$ preserving the spectrum of the matrix $(iM)$, i.e., preserving 
$\o_1,\ldots,\o_D$ up to the signs. Transformations $R\in SO(2D)$ with 
$\det R=1$ keep the system in the same phase. There is a discrete $Z_2$ 
transformation that changes the sign of one eigenvalue, we choose $\o_D$ for 
definiteness. We represent this transformation using the 
flip matrix $F$, $F_{ii}=1, i=1,\ldots,2(D-1),\;F_{2D-1,2D}=F_{2D,2D-1}=1$, 
and all 
other matrix elements zero.  There is a simple example of this duality 
transformation that connects the two phases, obtained by 
choosing $R'=FRF$ (see Fig.1):
\baa\label{example}
&&\o_D=-\o_D',\;\o_i=\o_i', i=1,\ldots,D-1,\nn
&&FMF=M',\;\prod \o_i=-\prod\o_i'.\nonumber\ea
In general, duality is characterized by $|\o_i|=|\o_i'|, \forall i$ and 
$\kappa=-\kappa'$.

The matrix $M$ can be brought to the $J_{\o}$ form by the orthogonal 
transformation $R$,
see Eq.(\ref{r2}). This $R\in SO(2D)$ matrix 
is unique up to orthogonal transformations that preserve $M$.
For a fixed values $\{\o_1,\ldots,\o_D\}$, the number of
parameters of the matrix $M$ is the same as  the number of
parameters of $R$.
The group of orthogonal transformations keeping $M$ invariant,
$SO(2D)\cap G(M)$, is 
isomorphic to $[U(1)]^D$ in the generic case. 
Using this freedom we can fix
$M_{2i-1,2i}=1,\forall i$ or we can put  
$M_{2D-1,2D}=-1$.
So, using the symmetry we reduce the number 
of continuous phase-space parameters to $2D(D-1)$.

For a special choice of phase-space parameters $M_{ij}$, we can enlarge the
symmetry group $[U(1)]^D$. 
The symmetries are characterized by degeneracy of eigenvalues $|\o_i|$.
If  $k_1,\ldots,k_{\alpha}$ are frequencies of appearence of $|\o_1|,\ldots,
|\o_{\alpha}|$
in the spectrum of the matrix $iM$, then the symmetry  group $SO(2D)\cap G(M)
\sim
U(k_1)\times\cdots\times U(k_{\alpha})$, where $\sum k=D$.
Obviously, the largest symmetry group is $U(D)$.
The sign of the product of eigenvalues determines the phase and the degeneracy 
among $|\o_i|$'s determines the complete symmetry structure of phase space.
In this way, we classify \nc \ spaces according to
$\{\o_1,\o_2,\ldots,\o_D\}$.

Figure 1 is, of course, valid in any number of dimensions, and 
we can construct corresponding transformations in a
way analogous to the two-dimensional case.

After defining the \h \, we can also discuss the group of 
linear transformations $G(H)\subset Gl(2D,\Rb)$ that keep \h \ invariant.
For the \nc \ harmonic oscillator, this group is $O(2D)$.
The degenerate energy levels for a given \h \ are described by a 
set of orthogonal eigenstates transforming according to an  
irreducible representation of the dynamical symmetry group.
The dynamical symmetry group $G(H,M)$ is a group of all transformations 
preserving both, $M$ and the \h \, i.e., $G(H,M)=G(H)\cap G(M)$.
For the fixed \h \, the dynamical symmetry depends on $M$, so, by changing 
the parameters of the matrix $M$ we can change $G(H,M)$ from $G_{min}(H,M)$ 
to $G_{max}(H,M)$.
For the \nc \ harmonic oscillator, the minimal dynamical symmetry group 
is $[U(1)]^D$, and the maximal symmetry is $U(D)$.
Note, however, that after fixing both the \h \ and $M$, all systems
connected to $(H,M)$ by linear transformations will have 
dynamical symmetry groups isomorphic to each other.

Hence, different choices of $M$ correspond to different dynamical 
symmetry. This can be viewed as a new mechanism of symmetry breaking
with the origin in (phase)space structure.  There are possible applications
to bound states in atomic, nuclear and particle physics. From the 
symmetry-breaking effects in these
systems one can, in principle, extract 
upper limits on the \nc \ parameters.

\section{Harmonic oscillator - an example}

In order to illustrate general claims from the preceding sections,
we  choose a 
simple harmonic oscillator in two dimensions as an example.
The  $O(4)$ invariant \h \ in this case is
\bee\label{ho}
H=\hlf\sum_{i=1}^4 U_i^2.\eeq
The constants $\hbar$, $m$ and $\o$ are absorbed in phase-space variables.
We parametrize the matrix $M$ by four parameters:
\baa\label{mho}
M=\left(\begin{array}{cccc}
 0&1&\theta&\phi_1\\
 -1&0&\phi_2&B\\
 -\theta&-\phi_2&0&1\\
 -\phi_1&-B&-1&0\end{array}\right).\ea
The eigenvalues of the matrix $iM$ are 
\baa\label{opho}
\o_{1,2}&=&\frac{1}{2}\sqrt{(\theta-B)^2+(\phi_1+\phi_2)^2+4}\nn
&\pm& \frac{1}{2}\sqrt{(\theta+B)^2+(\phi_1-\phi_2)^2},\ea
and the spectrum of the \h \ (\ref{ho}) is $E=\o_{1}(n_1+1/2)+|\o_{2}|
(n_2+1/2)$ \cite{nairp}, see Eq.(\ref{trans}) bellow.
If the product of eigenvalues is positive, we are in phase I, 
and if negative in phase II. The frequency $\o_1$ is always positive,
and $\o_2$ changes the sign in phase II.
Duality relations between the two phases are obtained by demanding that 
physical systems in both phases have the same energy spectrum.
In the simple case 
$\phi_1=\phi_2=0$, we have one-to-one correspondence between $(\theta,B)$ and 
$(\theta',B')$
\baa\label{dualho}
\theta &=& \frac{1}{2}\left[\sqrt{(\theta'-B')^2+4}
+\sqrt{(\theta'+B')^2-4}\;\right],\nn
B &=& \frac{1}{2}\left[\sqrt{(\theta'-B')^2+4}-
\sqrt{(\theta'+B')^2-4}\;\right],\ea
and
\baa\label{dualho2}
\theta' &=& \frac{1}{2}\left[\sqrt{(\theta-B)^2+4}
+\sqrt{(\theta+B)^2-4}\;\right],\nn
B' &=& \frac{1}{2}\left[\sqrt{(\theta-B)^2+4}-
\sqrt{(\theta+B)^2-4}\;\right].\ea
A comment is  in order. Notice that 
relations (\ref{dualho}) and (\ref{dualho2}) are 
valid for $|\theta+B|>2$ and $|\theta'+B'|>2$, respectively.
This is the sole consequence of the oversimplified parametrization
$\phi_1=\phi_2=0$. For every point in parameter space there
exists a dual point, we just have to allow for the most general 
parametrization
 of $M$.
Finally, from $\o_1\o_2=-\o'_1\o'_2$,  we obtain
\bee\label{dualp}
1-\theta B=\theta' B'-1.\eeq
This condition is necessary but not sufficient in order to have 
energy spectra in two phases identical.
A special case ($\theta=\theta'$)
of this relation was obtained in Ref.\cite{nairp},
by considering the limit from the fuzzy sphere to the plane, for the
Landau problem.

Although the systems depicted in Fig. 1 are physically distinct,
the dynamical symmetry groups are all isomorphic to each other.
 At every point in Fig. 1 the generic symmetry
($\o_1\neq |\o_2|$) is $U(1)\times U(1)$.  We have only one
quadratic symmetry generator, in addition to the  \h \
\bee\label{G}
{\cal G}=\sum_{i,j}C_{ij}U_iU_j,\;\;[{\cal G},H]=0.\nonumber\eeq
The matrix $C$ is symmetric, commutes with $M$, $[C,M]=0$,
and we can always choose ${\rm Tr} C=0$.
Then, $C^2$ is proportional to the identity matrix.
Namely, the $C^0$ matrix for the system $(U^0,J_{\o})$ is 
$C^0\sim {\rm diag}(1,1-1,-1)$. Using the $R$ transformation $U=RU^0$
we obtain $C=RC^0R^T$ implying $C^2\sim \Ii_{4\times 4}$.
For the matrix $M$ (\ref{mho}), the  generator
commuting with the \h \   (\ref{ho}) is
\baa\label{opci}
{\cal G}&=&\frac{1}{1-\theta B+\phi_1\phi_2}\left\{ (B+\theta)(X_1P_2-X_2P_1)
\right.\nn &-&
\frac{1}{4}(\theta^2-B^2+\phi_1^2-\phi_2^2)X_1^2
-\frac{1}{4}(\theta^2-B^2-\phi_1^2+\phi_2^2)X_2^2\nn &+&\frac{1}{4}(\theta^2-B^2+
\phi_1^2-\phi_2^2)P_1^2
+\frac{1}{4}(\theta^2-B^2-\phi_1^2+\phi_2^2)P_2^2\nn &-&
(\phi_1-\phi_2)(X_1X_2+P_1P_2)
-(B\phi_1+\theta\phi_2)X_1P_1\nn &-&\left.(B\phi_2+\theta\phi_1)X_2P_2
\right\}.
\ea
One is tempted to call this generator the angular momentum, but
this requires caution, as we have already discussed. 
For example, in the system
$(H(RU^0),J_{\o})$
we cannot construct the angular momentum because $[E,J_{\o}]\neq 0$. However,
the symmetry generator for this system is
$${\cal G}_0=\frac{1}{2\sqrt{\o_1|\o_2|}}\left(X_1^{0\;2}+P_1^{0\;2}-X_2^{0\;2}
-P_2^{0\;2}\right).$$

There are special points in parameter space of enhanced symmetry.
In the special
case $\o_1=\o_2$ (in phase I), we have the $U(2)$ symmetry group.
In this case  $\hbar_1=\hbar_2=1$, $B=-\theta$ and 
$\phi_1=\phi_2=\phi$ and we can  construct
three  generators of dynamical symmetry
satisfying the 
$SU(2)$ algebra $[L_i,L_j]=i\varepsilon_{ijk}L_k,\;i,j,k=1,2,3$:
\baa\label{su2}
L_1&=&\frac{1}{1+\theta^2+\phi^2}\left[X_1P_2-X_2P_1
-\phi(X_1P_1+X_2P_2)\right.\nn &-&\left.\frac{\theta}{2}(X_1^2+
X_2^2-P_1^2-P_2^2)\right],\nn
L_2&=&\frac{1}{1+\theta^2+\phi^2}\left[-P_1P_2-X_1X_2+\theta(X_1P_1-X_2P_2)
\right.\nn &+&\left.
\frac{\phi}{2}(X_2^2-X_1^2+P_1^2-P_2^2)\right],\nn
L_3&=&\frac{1}{1+\theta^2+\phi^2}\left[\frac{1}{2}(X_1^2-X_2^2+P_1^2-P_2^2)
\right.\nn &+&\left.\theta(X_1P_2+X_2P_1)-\phi(X_1X_2-P_1P_2)\right].\ea
The dual point with $\o_1=-\o_2$, with $SU(2)$ symmetry in phase II, 
is obtained with
$B=\theta,\phi_1=-\phi_2=\phi,\hbar_1=-\hbar_2=1$.
We wish to emphasize that $SU(2)$ symmetry exists only for a
special choice of parameters, and is not a dynamical symmetry of the 
\h \ in the generic case (in contrast to the claims in Ref.\cite{sma}).

The transformations ${\cal L},{\cal L}_0,S,D,R$ connecting
different representations (see Fig. 1) of the harmonic 
oscillator on the \nc \ plane were discussed in the preceding 
section, and, partly, in the literature \cite{nairp,smyr,sma}.
Using the matrix ${\cal L}_0=RD$ we can transform the \h \ (\ref{ho}) into 
an ordinary QM system:
\baa\label{trans}
H(U)&=&H(RDu^0)=\frac{1}{2}{\cal L}_{ik}^0{\cal L}_{il}^0u_{k}^0u_l^0
\nn &=&
\frac{\o_1}{2}(u_1^{0\;2}+u_2^{0\;2})+\frac{|\o_2|}{2}(u_3^{0\;2}+u_4^{0\;2})
.\ea
Next,  we calculate matrix elements of observables
starting form ordinary harmonic oscillator observables:
$$\langle U_i\cdots U_k\rangle={\cal L}^0_{ij_1}\cdots{\cal L}^0_{kj_k}
\langle u_{j_1}^0\cdots u^0_{j_k}\rangle. $$
For quadratic observables in the ground state we use
$\langle u_i^{0\;2}\rangle=1/2,\;\langle u_1^0u_2^0\rangle=
\langle u_3^0u_4^0\rangle=i/2$, all others are zero.  
For a special case $\phi_1=\phi_2=0$, we use the matrix $R$ Eq.(\ref{rr}) 
to obtain 
\baa\label{mx}
\langle X_1^2\rangle=\langle X_2^2\rangle=\frac{1}{2}
\left[\o_1\cos^2{\varphi}+|\o_2|\sin^2{\varphi}\right],\nn
\langle P_1^2\rangle=\langle P_2^2\rangle=\frac{1}{2}
\left[\o_1\sin^2{\varphi}+|\o_2|\cos^2{\varphi}\right].\ea
These expressions are universal, i.e., they are 
valid in both phases and at the critical point.

Here, we would like to comment 
uncertainty relations following from commutation
rules which define the theory. 
In the simple case $\phi_1=\phi_2=0$, we have four nontrivial 
uncertainty relations $\Delta U_i\Delta U_j\geq |M_{ij}|/2$, 
i.e.,
\bee\label{unc1}
\langle X_a^2\rangle\langle P_a^2\rangle\geq \frac{1}{4},\;a=1,2,\eeq
\bee\label{unc2}
\langle X_1^2\rangle\langle X_2^2\rangle\geq \frac{\theta^2}{4},\;
\langle P_1^2\rangle\langle P_2^2\rangle\geq \frac{B^2}{4}.\eeq
We calculate the left-hand-side of relations (\ref{unc1},\ref{unc2}) in the 
ground state, 
using (\ref{mx}) and (\ref{cs}).
In phase I we can 
saturate the first  two relations (\ref{unc1}) 
for $\theta=B$.
In  phase II we can saturate
the other two relations (\ref{unc2})
for any $B$ and $\theta$.  
At the critical point 
$\theta B=1$ all four relations are saturated in the ground state.
In the special case in phase I, $B=0,\theta\neq 0$, none of the four 
uncertainty relations  are saturated, 
in agreement with  the  theorem valid for quantum mechanics
on the noncommutative plane  with $B=0$ \cite{un}. 
This short analysis also indicates that physics in different 
phases is qualitatively different and depends crucially on $M$.

An especially interesting physical system is the
Landau problem  in the \nc \ plane, defined by 
$H={\bf P}^2/2$ and the matrix $M$
\baa\label{mhoi}
M=\left(\begin{array}{cccc}
 0&1&\theta&0\\
  -1&0&0&B\\
   -\theta&0&0&1\\
    0&-B&-1&0\end{array}\right).\ea
This problem can be treated as a \nc \ harmonic oscillator $H=
{\bf P}^2/2+\o^2{\bf X}^2/2$, in the limit
when $\o\rightarrow 0 $. We simply define $U_1=\o X_1$, $U_3=\o X_2$ to 
obtain a new  matrix $M_{\o}$
\baa\label{mhai}
M_{\o}=\left(\begin{array}{cccc}
 0&\o&\o^2\theta&0\\
   -\o&0&0&B\\
      -\o^2\theta&0&0&\o\\
          0&-B&-\o&0\end{array}\right),\ea
with the determinant $\det M_{\o}=\o^4(1-\theta B)^2$.
We find the magnetic length (the minimum spatial extent of the wave-function 
in the ground state) in a universal form, valid
in both phases and at the critical point:
\bee\label{rx}
\langle X_1^2+X_2^2\rangle=\langle r^2\rangle=
\frac{
 |1-\theta B|+1+\o^2\theta^2}{\sqrt{(\o^2\theta-B)^2+4\o^2}}.\eeq
In the limit $\o\rightarrow 0 $, eigenvalues\footnote{In the "exotic" 
approach \cite{ph} $\o_1^{\rm ex}=\o_1/\kappa=1/\o_2, \;\o_2^{\rm ex}
=\o_2/\kappa=1/\o_1$, and in the limit $\o\to 0$ eigenvalues are 
$|\o_1^{\rm ex}|\to\infty,
\;\o_2^{\rm ex}\to 1/|B|$.} of the matrix $M_{\o}$ are 
$\o_1=|B|,\;\o_2=0$  and magnetic length is 
\baa
\langle r^2\rangle=\left\{\begin{array}{ccc}
\frac{2-\theta B}{|B|}&{\rm if}&\theta B<1\\
\theta' &{\rm if}&\theta'B'>1\end{array}\right.\ea
For $|B|=B'$ these two expressions are the same if the 
duality relation (\ref{dualp}) holds.

The above representation of the \nc \ Landau problem as a 
case of \nc \ harmonic oscillator with $\o\rightarrow 0$ can also
be  viewed as a \nc \ harmonic oscillator with $\tilde \o\neq 0$,
at the  critical point $\tilde \theta\tilde B=1$.
The connection between parameters is 
$\tilde\o^2\tilde\theta+1/\tilde\theta=B$.
If we insist on having the same magnetic length in both 
pictures, we can fix $\tilde\theta$ and $\tilde\o$.

However, these systems (the Landau problem with $\o=0$ 
and the 
harmonic oscillator with $\tilde \o\neq 0$) are not physically equivalent. 
They have just the same energy 
spectrum  and the same magnetic length if we choose so.
A simple way to see this is to consider uncertainty relations in 
phases I  and II for the Landau problem, and at the critical point for 
the harmonic oscillator.
Here we wish to emphasize once more that only system having 
equal  both the spectrum of the \h \ 
and the matrix of commutators $M$ are physically equivalent.

\section{Conclusion}

We have presented a unified approach to NCQM in terms of \nc \ coordinates 
and momenta in arbitrary dimensions and for arbitrary 
c-number  commutation relations. We have considered all representations 
of NCQM connected by linear transformations from $Gl(2D,\Rb)$ preserving the 
property that commutation relations remain independent of phase-space variables
and keeping the energy 
spectrum of the system fixed. Among these only representations connected by 
transformations preserving the commutation relations are physically 
equivalent.
We classify \nc \ spaces according to the eigenvalues of the matrix $iM$,
$\{\o_1,\o_2,\ldots,\o_D\}$. 
The sign of the 
product of eigenvalues determines the phase and the degeneracy
among $|\o_i|$'s determine the complete symmetry structure of phase space.
Since   orthogonal transformations keep the spectrum of the matrix
$iM$ fixed, 
they have been analysed in
detail. 
We have shown that for general $M$ the angular momentum operator 
in the usual sense might
not exist, and  we have given the condition for its existence.
An important result is that two physically distinct phases  exist in 
arbitrary dimensions and  that they are
connected by discrete duality transformations.

Besides the symmetry structure of space,  we have also discussed
the dynamical symmetry
of a physical system and proposed  a new mechanism for symmetry breaking,
originating from phase-space structure.

In our approach to symmetries, there is no physical principle what $H$ 
and $M$ we have to choose in terms of noncommuting variables $U$.
One way to test the idea of noncommutativity is to 
choose the \h \ as in ordinary quantum mechanics, and to search for 
(tiny) symmetry breaking effects induced by the phase-space structure $M$.
The opposite way \cite{esp} is to fix the dynamical symmetry structure as in 
ordinary quantum mechanics. In the latter case, the differences shoud appear
in matrix elements of observables and energy eigenstates.
Of course, one can choose a combination of both approaches.
Regardless of the approach,
noncommutativity offers a new explanation of symmetry breaking,
or change in probability amplitudes as a consequence of phase-space
(space-time) structure.

Our general approach enabled us to obtain new results  even
in the simplest case of two-dimensional harmonic oscillator. We expect
that we shall also obtain physically interesting results
in the $D=3$ case, currently under investigation.

\begin{acknowledgement}
We thank I. Dadi\'c and M. Milekovi\'c for useful discussions.
This work was supported by the Ministry of Science and Technology of the
Republic of Croatia under contract No. 0098003.
\end{acknowledgement}

\end{document}